\documentclass[useAMS,usenatbib,fleqn]{mn2e}

\usepackage{times}

\usepackage{amsmath}
\usepackage{amssymb}
\usepackage{upgreek}
\usepackage{graphicx}

\title[Flare Analysis]{Analysis of Methanol Maser Flares in G107.298+5.63 and S255-NIRS3}
\author[M. D. Gray, S. Etoka, A. and B. Pimpanuwat]
{M. D. Gray$^{1,2}$, S. Etoka$^{1}$ and B. Pimpanuwat$^{1,2}$\\
$^{1}$Jodrell Bank Centre for Astrophysics, Department of Physics and Astronomy, University of Manchester,
M13 9PL, UK\\
$^{2}$National Astronomical Research Institute of Thailand, 260 Moo 4, T. Donkaew, A. Maerim, Chiangmai 50180, Thailand.}
\begin{document}

\date{}

\pagerange{\pageref{firstpage}--\pageref{lastpage}} \pubyear{2020}

\maketitle

\label{firstpage}

\begin{abstract}
A 3D maser model has been used to perform an inverse problem on the light curves from three
high-amplitude maser flares, selected on the basis of contemporaneous infra-red observations.
Plots derived from the model recover the size of the maser cloud, and two parameters linked
to saturation, from three observational properties of the light curve. Recovered sizes are
consistent with independent interferometric measurements. Maser objects transition between
weak and moderate saturation during a flare.
\end{abstract}


\section{Introduction}
\label{s:intro}

\defcitealias{2020MNRAS.493.2472G}{Paper~1}
In a recent paper \citep{2020MNRAS.493.2472G} \citepalias{2020MNRAS.493.2472G}, we 
demonstrated that radiatively-driven maser flares may be usefully
characterised by three statistics derived from their light curves: the variability index, the
duty cycle, and the maximum flux density achieved during the flare. The variability index is
defined here as the maximum flux density divided by the minimum, if the source is periodic, or
the maximum flux density divided by the pre-flare quiescent level for an aperiodic source.
The duty cycle is defined as the fraction of a characteristic time for which the maser flux
density is above half the peak value; the characteristic time is the flare period for periodic
sources, and the time for which the flare is considered active for aperiodic sources.

The duty cycle is a particularly useful statistic because it provides a clear distinction
between flares that are driven by variation in the pumping radiation and flares that are
driven by variations in the background radiation. Maser flares of the former type rarely
have a duty cycle that exceeds that of the driving function, that is the light curve of
the infra-red (IR) pumping radiation, whilst the latter type almost invariably have a duty
cycle greater than that of the driving function of the radio background, at least for flares
of significant variability index ($>$1.5).

In the present work, guided by observational constraints on the duty cycle and variability
index, we provide fits to spectral components with flares in two star-forming
region sources: G107.298+5.63 and S255-NIRS3. The former source exhibits periodic flares in the
6.7-GHz maser transition of methanol, whilst the latter source is an example of an accretion
burst source that is aperiodic with extreme peak flux densities ($>$1000\,Jy) in some
spectral components. The period of the methanol masers in G107.298+5.63 is
34.6\,d, an interval that applies also to 22-GHz H$_2$O masers in the source \citep{2016MNRAS.459L..56S}.
Perhaps the most plausible physical model for the overall G107.298+5.63 flare mechanism is
a colliding-wind binary \citep{2018IAUS..336...37S}.
 G107.298+5.63 and S255-NIRS3 were chosen because they are examples of a currently
very select group that have IR light curves measured contemporaneously with the maser data.
In G107.298+5.63, we consider the spectral component at -7.4\,km\,s$^{-1}$ and in S255-NIRS3,
the components at 6.42 and 5.83\,km\,s$^{-1}$ (see Fig.~3 of \citealt{2020PASJ...72....4U}).

\section{Model and Observational Data}
\label{s:model}

The model used here is as described in \citetalias{2020MNRAS.493.2472G},
and is based on earlier versions developed in \citet{2018MNRAS.477.2628G} and \citet{2019MNRAS.486.4216G}.
Data used in \citetalias{2020MNRAS.493.2472G} was supplemented 
for the present work by a series of models with a shell-like
distribution of the unsaturated inversion with a distribution $\propto r^2$, where $r$ is the radial
distance from the domain centre. Data from the model with the shell distribution were used to 
estimate modelling uncertainties. All model domains were prolate objects with a distortion
factor of $\Gamma = 0.6$ (see \citet{2019MNRAS.486.4216G} for definition), and viewed from the optimum direction.
This combination of source shape and viewing direction were found to give the largest flux densities
for a source of given volume and mean unsaturated inversion. Viewing from
a random orientation reduces the peak flux density for this type of
object by approximately two orders of magnitude (see Fig.~15 of \citetalias{2020MNRAS.493.2472G}).

A convenient way to present data from a large number of models is to present one statistic, the
maximum flux density, as a colour palette, and the other two statistics via sets of contours, all on
the same set of axes. These axes show the initial optical depth parameter of the model on the
$x$ axis, noting that the depth parameter does not change in variable background models, and the
change in the driving function on the $y$ axis. A single point in the $xy$-plane of a diagram of this
type involves computing an entire maser light curve, even though only the three statistics derived
from it are plotted. When a fit has been established, it is then straightforward to recover the
complete maser light curve at that $xy$ position.

\begin{figure}
  \includegraphics[bb=60 50 560 435, scale=0.47,angle=0]{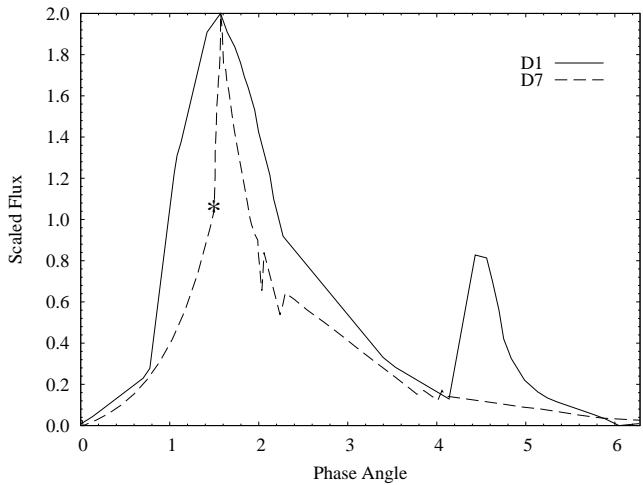}
  \caption{Driving functions $D_1$ and $D_7$. Both functions are scaled to unit amplitude and the maximum 
placed at $\pi/2$\,radians. The symbol `*' in $D7$ marks the start of the observational data. Points to
the left of this marker follow an exponential added by the present authors.}
\label{f:drivers}
\end{figure}

The observational data that we use as driving functions in this work are, firstly, the \textit{NEOWISE} data from
G107.298+5.63 \citep{2018IAUS..336...37S}. The digitized form of this IR light curve is referred to
as function $D1$ in \citetalias{2020MNRAS.493.2472G}, and we keep this
nomenclature here. Secondly, we use the IR light
curve in the K$_s$ band from S255-NIRS3 \citep{2020PASJ...72....4U}. This light curve is incomplete
at early times, and we have completed it by fitting an exponential function that links the earliest
epoch in Figure~3 of \citet{2020PASJ...72....4U} with the first observational data point. This
latter function was not used quantitively in \citetalias{2020MNRAS.493.2472G}, but
is named $D7$ here to follow on from the highest numbered function in \citetalias{2020MNRAS.493.2472G}. Both driving
functions are plotted in a normalized form in Fig.~\ref{f:drivers} with maximum light at a phase
angle of $\pi/2$. Digitization tasks were carried out using the WebPlotDigitizer tool, by 
Ankit Rohatgi\footnote{https://automeris.io/WebPlotDigitizer}.

\section{Analysis}
\label{s:analysis}

\begin{figure}
  \includegraphics[bb=70 70 620 260, scale=0.75,angle=0]{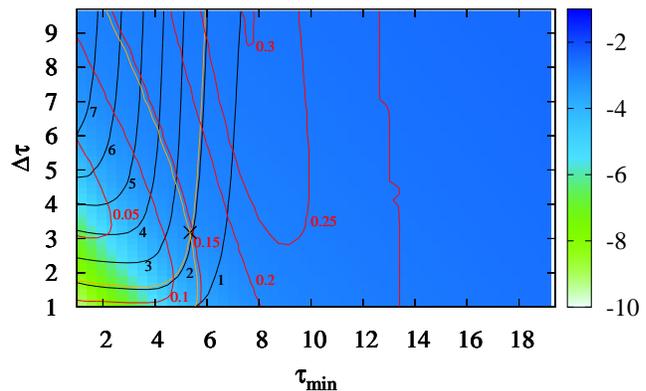}
  \caption{The maximum flux density achieved (represented by the colour scale), the base-10
logarithm of the variability index (black contours) and duty cycle of the maser (red contours)
as functions of the maser optical depth parameter at minimum light ($x$ axis) and the change
in this parameter in moving from minimum to flare maximum ($y$ axis). One contour from each
set is plotted in orange for the observationally derived variability index and duty cycle, with
the intercept marked by a black $\times$ symbol. This figure was constructed
using the driving function D1, making it specific to G107.298+5.63.}
\label{f:G107}
\end{figure}

We consider initially IR pumping. In this case, a change in optical depth parameter, as a proxy
for the unsaturated maser inversion, follows the driving function. We use function $D1$ for
the G107.298+5.63 flare and function $D7$ for S255-NIRS3. In our analysis of G107.298+5.63, the
principal difference between the present work and our preliminary analysis in \citetalias{2020MNRAS.493.2472G}
is that we now use $D1$ to construct the fitting plot, rather than a sinusoid, so that our new
plot in Fig.~\ref{f:G107} is source-specific. As in \citetalias{2020MNRAS.493.2472G} we use, for the
parameters of the maser response in the -7.4\,km\,s$^{-1}$ feature, a duty cycle of 0.143 and
the quoted variability index of 120 \citep{2016MNRAS.459L..56S}. We then read off from Fig.~\ref{f:G107}
the intercept of the orange contours at 0.143 and $\log_{10}(120)=2.08$ to recover
the original optical depth parameter of the maser and the change in depth needed to generate the
flare. The result is $(\tau_{min},\Delta \tau) = (5.14,3.15)$, marked with the $\times$ symbol
in Fig.~\ref{f:G107}, and these figures correspond to an oscillation between
conditions of weak and moderate saturation. Of course, there are significant modelling 
uncertainties attached to this fit, and to those calculated similarly below. We estimate the
size of these uncertainties in Section~\ref{s:discuss}.
The third parameter, the maximum flux density
achieved, is an additional constraint. At the intercept point given above, the scaled flux density
has a value of $f_\nu = 6.23\times 10^{-4}$ from the colour scale in Fig.~\ref{f:G107}. With an observational
flux density in Jy, and a distance to the source, we may calculate the required size of the maser
source from eq.(A6) of \citetalias{2020MNRAS.493.2472G}.
Inserting our dimensionless flux density, the observational counterpart of 57.7\,Jy,
a loss rate in the 6.7-GHz transition of 0.79\,Hz, as
in \citetalias{2020MNRAS.493.2472G}, and a source
distance of 0.76\,kpc \citep{2008PASJ...60..961H}, the model and measured flux densities 
are consistent if the maser object has a size of about 4.4\,AU. We show the observed maser light
curve and the light curve calculated at our fit position in the upper panel of Fig.~\ref{f:triple}.
The derived size of the maser object is consitent with \textit{EVN} observations, provided that
the sizes of individual maser objects are significantly smaller than an
observed region 30-80\,AU in extent \citep{2016MNRAS.459L..56S}, that contains several such
objects. 

\begin{figure}
  \includegraphics[bb=70 70 620 260, scale=0.75,angle=0]{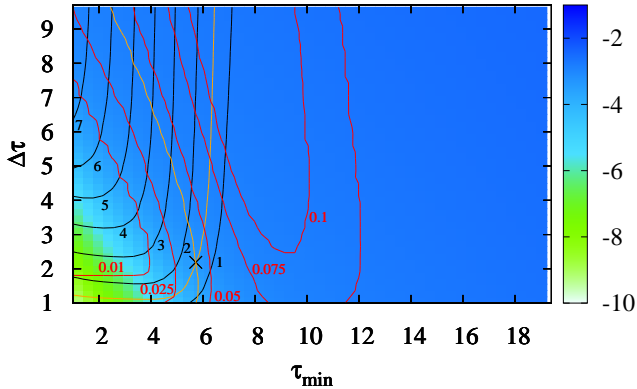}
  \caption{As for Fig.~\ref{f:G107}, but prepared with the $D7$ driving function,
specific to the S255-NIRS3 source. The black cross marks the 
variable pumping fit position for
the -5.83\,km\,s$^{-1}$ feature.}
\label{f:S255}
\end{figure}

In S255-NIRS3, there are two spectral features that might be considered problematic for analysis in
view of their large flux density and/or variability index. We consider first the feature at
5.83\,km\,s$^{-1}$. For this feature, the maser flux density is above half
the peak value for 45 days out of the 1100 flaring days, corresponding to 
a duty cycle of 0.041, and its variability
index is quoted as 27 \citep{2020PASJ...72....4U}. We attempt to fit these parameters with a new
plot, Fig.~\ref{f:S255}, that differs from Fig.~\ref{f:G107} in being prepared with the IR driving
function $D7$, instead of $D1$, making Fig.~\ref{f:S255} specific to S255-NIRS3. Using the same fitting
procedure applied above, we find that the 0.041 orange contour meets its variability index
counterpart, delineating $\log_{10}(27)=1.43$,
at $(\tau_{min},\Delta \tau) = (5.72,2.19)$. The dimensionless flux density at this point
is $f_\nu = 4.22\times 10^{-4}$. To make this consistent with an observed peak flux density of 1632\,Jy,
with S255-NIRS3 at a distance of 1.78\,kpc \citep{2016MNRAS.460..283B}, requires a maser object 
with a size of 67\,AU. We show the observed maser light curve and our light curve at the fitting
point in the middle panel of Fig.~\ref{f:triple}.

\begin{figure}
  \includegraphics[bb=55 70 620 260, scale=0.71,angle=0]{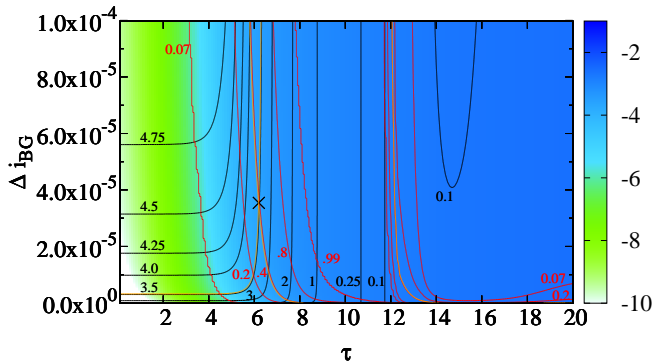}
  \caption{As for Fig.~\ref{f:G107}, but prepared with $D7$ driving function,
and the $y$ axis variable is now the change in the intensity of background radiation. Use of
the $D7$ function makes the plot specific to the S255-NIRS3 source. The cross marks the
fit position for the -6.42\,km\,s$^{-1}$ feature.}
\label{f:bg3}
\end{figure}

\begin{figure}
  \includegraphics[bb=380 90 620 790, scale=0.58,angle=0]{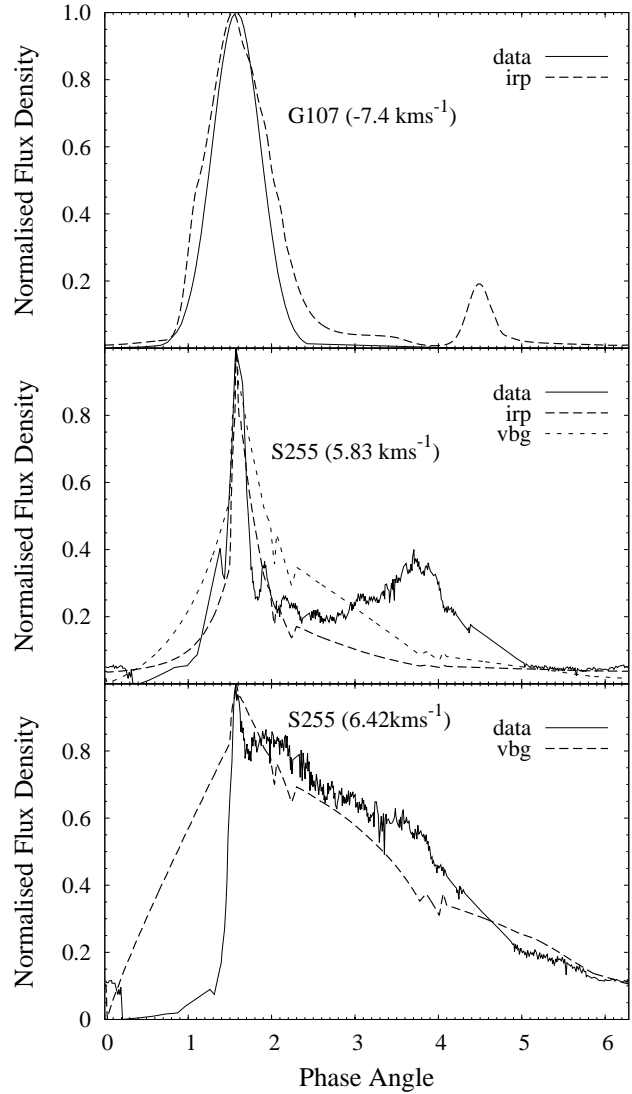}
  \caption{Our fits based on two model parameters compared with observational maser flare
data for the three maser light curves analysed. Fits using the IR pumping mechanism are
marked `irp'; those using the variable background are marked `vbg'.
Observational data for the S255-NIRS3 features
(middle and bottom panels) were digitized from graphs in \citet{2020PASJ...72....4U}. Data
for G107.298+5.63 (top panel) follow the best-fit gaussian in \citet{2014PASJ...66...78F}, since
no light curve is plotted for the -7.4\,km\,s$^{-1}$ feature in \citet{2016MNRAS.459L..56S}, and
only a copy of the best-fit gaussian is plotted in \citet{2018IAUS..336...37S}. In all cases,
the functions are plotted over a phase-angle range of 0-$2\pi$ with the flux densities normalised
to the values at maximum light. This amounts to choosing the cloud size that makes the model
flux density agree exactly with the observed value in Jy (see text above). Times corresponding
to a phase angle of $2\pi$ are 34.6\,d (the maser period) in the top panel and 1100\,d in the
other two panels. 
}
\label{f:triple}
\end{figure}

The maser feature at 6.42\,km\,s$^{-1}$ in S255-NIRS3 has no fit point in Fig.~\ref{f:S255}: its
duty cycle of 0.39 is too large to be consistent with its variability index of 3400 (the base 10 logarithm
of this number is 3.531). On these grounds, we reject an IR pumping mechanism for this component and
seek instead a fit to a variable background mechanism. There is an immediate problem with this approach
because the $D7$ driving function is based on IR observations in the K$_s$ band, not on the radio continuum
background at 6.7\,GHz. However, in the absence of any contemporaneous radio background data, we still
use the $D7$ driving function to prepare Fig.~\ref{f:bg3} in the knowledge that this functional form
could be quite wrong. The $y$ axis in Fig.~\ref{f:bg3} is now the change in the intensity of the
background radiation; the optical depth parameter, once chosen, is now a constant during the flare.
With $D7$ as the driving function, it becomes possible to find a fit for the 6.42\,km\,s$^{-1}$ feature.
The intercept of the 0.39 duty-cycle contour and the $\log_{10}(3400)=3.53$ variability contour 
(both orange in Fig.~\ref{f:bg3}) occurs at 
$(\tau,\Delta i_{BG})=(6.20,3.56\times 10^{-5})$, noting that $i_{BG}$ is measured in units of the saturation
intensity of the maser transition.

The scaled flux density at the position of the fit in Fig.~\ref{f:bg3} is $2.432 \times 10^{-4}$, and
this figure, to be consistent with an observed flux density of 1705\,Jy (see 
Table~1 of \citealt{2020PASJ...72....4U}), requires a cloud size
of 89.6\,AU. This is a similar size to that required for the feature at 5.83\,km\,s$^{-1}$, even though the
pumping mechanisms are probably different, and is a consequence of the similar observed peak flux densities and
model flux densities achieved for the two features. The maser light curve and our model counterpart
are shown in the bottom panel of Fig.~\ref{f:triple}.
The required cloud size is probably consistent with results from VLBI observations: Observations
of S255-NIRS3 with the \textit{EVN} during the burst resolved out over 90 per cent of the single-dish
flux density with a beam of $3 \times 5$\,milliarcsec \citep{2018A&A...617A..80S}. 
At 1.78\,kpc, this implies that most of the
maser emission arises from a structure, or structures, significantly larger than 7\,AU. Lower resolution 
\textit{JVLA} observations suggest the maser structures could be up to 430\,AU in size
\citep{2017A&A...600L...8M}. Our values of 60-90\,AU, required for
consistency between model and observational flux densities in S255-NIRS3, therefore fall neatly into the
range of scales associated with the majority of the maser flux density.

If there is a requirement that the same (variable background) mechanism should be used for the 5.83\,km\,s$^{-1}$
feature, then an approximate fit can be made from Fig.~\ref{f:bg3}. In this case we demand a model flux
density (1.02$\times$10$^{-5}$) corresponding to the possible 430\,AU structures in S255-NIRS3, but compromise
on the duty cycle (0.080) and variability index (624) that are found at 
$(\tau, \Delta i_{BG}) = (5.7,7.0\times 10^{-7})$. The fit with these parameters is shown as the curve
marked `vbg' in Fig.~\ref{f:triple}.

\section{Discussion}
\label{s:discuss}

We have demonstrated that our global model of radiation-driven maser flares can be used to fit
a variety of maser light curves, including `giant flares', with peak flux densities in excess
of 1000\,Jy for sources at distances of order 1\,kpc. No novel mechanisms, such as, for
example, superradiant emission \citep{2017SciA....3E1858R} are necessarily required. The
variability index and duty cycle parameters usually exclude one
or other of the IR pumping and variable background mechanisms. We have modelled the 5.83\,km\,s$^{-1}$
feature with both mechanisms, but only the IR pumping fit can accommodate all the observational constraints.
The fitting procedure is quite simple,
involving only three observational parameters of the maser flare that are used to recover
three model variables related to the amplification and saturation of the maser. The 3D nature
of the model allows for a reasonably strong consistency check by calculating the required size
of the maser object needed to equate the observed and modelled flux densities.

We have attempted to estimate uncertainties in our fits by calclating additional contour
intercepts for a different model in which the unsaturated inversions rise as the square of
the distance from the centre of the model cloud. We refer to this as the shell distribution.
This distribution is arguably a better representation of the unsaturated inversion
than a uniform model in the case of pumping by an optically thick
IR transition. In the same order as presented above, the alternative fits are
$(\tau_{min},\Delta \tau) = (4.65,3.91)$ with a flux density of 1.1$\times$10$^{-3}$
for G107.298+5.63, and
$(\tau_{min},\Delta \tau) = (5.00,2.75)$ with a flux density of 5.7$\times$10$^{-4}$ for
the 5.83\,km\,s$^{-1}$ feature of S255-NIRS3. No fit could be found for the variable background
model and shell profile in the case of the 6.42\,km\,s$^{-1}$ feature. The closest approach if the
correct duty cycle is required is $(\tau,\Delta i_{BG}) = (5.0,8\times10^{-5})$, where the
variability index is 2150. However, in the case of a variable background, the
reason for the adoption of the shell distribution of unsaturated inversion no longer applies.

Our estimation of uncertainties suggests that $\tau_{min}$, or $\tau$ for the background model, is
uncertain by approximately $\pm$1. For the IR pumped fits, the uncertainty in $\Delta \tau = 0.66$
averaged over both fits. The change in background is considerably more uncertain as no fit
was found for the shell model, and because of the almost vertical contours close to the fit
position in Fig.~\ref{f:bg3}.
Maser parameters recovered from the analysis show that, in all three cases modelled here, the
pre-flare, or minimum light, maser objects have only modest saturation, with a narrow range
of 5.14-6.20 in optical depth parameter within the
uniform model. An uncertainty of $\pm 1$ in $\tau_{min}$
does not change this conclusion.
At flare maximum, flux densities achieved by the
flaring objects are significant, but still a factor of 5-10 lower than those corresponding to a strongly-saturated
maser. Flux densities derived from the shell-model fits are approximately a factor of two
larger than for the uniform model, dividing required cloud sizes by a factor of $\sim \sqrt{2}$.

A glance at Fig.~\ref{f:triple} suggests that the exponential used to fill in the missing
IR data at early times in S255-NIRS3 is not particularly good, and that the real function is
something even steeper. We note that our model cannot reproduce secondary peaks, such as that
visible in the middle panel of Fig.~\ref{f:triple} unless such a peak is also present in the
driving function.
The use of the $D7$ IR function in the case of the variable background 
fit is highly speculative, but a fit could be obtained for the uniform cloud model. It should
be noted that the difference in background over the flare is important, rather than the ratio,
so results are dominated by the largest background levels used: a background range of
$10^{-6}-10^{-4}$ would therefore produce an almost identical Fig.~\ref{f:bg3} to the one
plotted, which uses $10^{-9}-10^{-4}$. For a frequency in the Rayleigh-Jeans limit, and parameters
for the 6.7-GHz transition of methanol from \citetalias{2020MNRAS.493.2472G}, the background
intensity at the fit position in Fig.~\ref{f:bg3} corresponds to bathing the maser cloud in
unattenuated, and geometrically
undiluted, black-body radiation at a temperature of 3080\,K.

Future models could include a core-halo density structure, allowing us to address the possible
presence of objects with scales of order 1-500\,AU and the issue of missing flux in VLBA observations
when compared with single-dish spectra.

\section*{Acknowledgments}

MDG and SE acknowledge funding from the UK Science and Technology Facilities
Council (STFC) as part of the consolidated grant ST/P000649/1 to the Jodrell Bank
Centre for Astrophysics at the University of Manchester. MDG acknowledges
financial support from the National Astronomical Research Institute of Thailand (NARIT)
whilst on sabbatical at their HQ in Chiang Mai, Thailand.
This work was performed, in part, using the DiRAC Data Intensive service at Leicester, operated 
by the University of Leicester IT Services, which forms part of the STFC 
DiRAC HPC Facility (www.dirac.ac.uk). The equipment was funded by BEIS capital 
funding via STFC capital grants ST/K000373/1 and ST/R002363/1 and 
STFC DiRAC Operations grant ST/R001014/1. DiRAC is part of the National e-Infrastructure.
Data used in this work was generated under DiRAC award dp124.

\section*{Data Availability Statement}

The data underlying this article will be shared on reasonable request to the corresponding author.

\bibliographystyle{mn2e}
\bibliography{let1}

\end{document}